\documentclass[aps,prb,showkeys,showpacs,floats]{revtex4}
\usepackage{epsfig}
\usepackage{amsmath}
\usepackage{amssymb}
\newcommand{\bd}{\begin{displaymath}}
\newcommand{\ed}{\end{displaymath}}
\newcommand{\be}{\begin{equation}}
\newcommand{\ee}{\end{equation}}
\newcommand{\bea}{\begin{eqnarray}}
\newcommand{\eea}{\end{eqnarray}}
\newcommand{\bse}{\begin{subequations}}
\newcommand{\ese}{\end{subequations}}
\newcommand{\ba}{\begin{array}}
\newcommand{\ea}{\end{array}}
\newcommand{\nn}{\nonumber}

\newcommand{\gam}{\gamma}
\newcommand{\al}{\alpha}
\newcommand{\sig}{\sigma}

\newcommand{\noi}{\noindent}
\newcommand{\eps}{\epsilon}
\newcommand{\lam}{\lambda}
\newcommand{\ua}{\uparrow}
\newcommand{\da}{\downarrow}

\newcommand{\ra}{\rangle}
\newcommand{\la}{\langle}
\begin{document}

\title{\bf An analytical decomposition protocol for optimal implementation of two-qubit entangling gates}

\author{
M. Blaauboer and R.L. de Visser
}

\affiliation{Kavli Institute of Nanoscience, Delft University of Technology,
Lorentzweg 1, 2628 CJ Delft, The Netherlands
}
\date{\today}

\begin{abstract}
This paper addresses the question how to implement a desired two-qubit gate $U$ 
using a given tunable two-qubit entangling interaction ${\cal H}_{int}(t)$. 
We present a general method which is based on the $K_1AK_2$ decomposition of unitary matrices
$\in SU(4)$ to calculate the smallest number of two-qubit 
gates $U_{int}(t)$ [based on ${\cal H}_{int}(t)$] and single-qubit rotations, and the explicit 
sequence of these operations that are required to implement $U$.  
We illustrate our protocol by calculating the implementation of (1) the transformation 
from standard basis to Bell basis, (2) the CNOT gate, and (3) the quantum Fourier transform
for two kinds of interaction - Heisenberg exchange
interaction and quantum inductive coupling - and discuss the relevance 
of our results for solid-state qubits. 
\end{abstract}

\maketitle

\section{Introduction}

How to implement two-qubit quantum circuits in an optimal way, by which we mean here that they require 
a minimum amount of operations, has become a topic of active interest over 
the last years. Typically, the central question asked is how to perform 
an arbitrary operation $\in SU(4)$ with the least number of two-qubit gates such as 
the CNOT (controlled NOT) gate. This has led to the discovery that any two-qubit 
operation can be implemented using three CNOT gates and local gates in between~\cite{vida04}. 
Alternatively, two-qubit operations can be constructed by three applications of the
(SWAP)$^{\al}$ gate\cite{fan05}, with $0<\al \leq 1$, or by two applications of the 
B-gate~\cite{zhan04}. From a practical point of view, it is essential to perform 
quantum operations as fast as possible, before decoherence (due to interactions 
with the environment) sets in. This is of particular importance for solid-state qubits, 
which in general interact strongly with their environment. For a specific physical 
system, however, efficient implementation of quantum operations 
does not necessarily coincide with minimizing the number of CNOT gates, 
if CNOT cannot be directly generated by the given two-qubit Hamiltonian. Rather, efficient 
construction of quantum operations requires minimizing the number of two-qubit gates which are 
based on the (entangling) two-qubit interaction that is naturally available in the system, as well as
minimizing the number of single qubit operations (which in practice generally also take a 
finite amount of time to implement~\cite{time}).

In this paper, we present a general, explicit and self-contained analytical protocol 
for calculating the implementation of a desired  
two-qubit quantum gate $U\in SU(4)$ as a sequence of single-qubit rotations and 
two-qubit gates $U_{int}(t)$, where $U_{int}(t)$ is based on the two-qubit interaction ${\cal H}_{int}(t)$
that is naturally available in a given system, $U_{int}(t) \equiv \exp[-\frac{i}{\hbar} 
\int_0^t dt^{'}\, {\cal H}_{int}(t^{'})]$.
We assume that $U_{int}(t)$ has some entangling capacity, which is a necessary requirement if $U$ itself is
an entangling gate. We also assume that $U_{int}(t)$ is a {\it tunable} gate, i.e. that one can switch 
the interaction ${\cal H}_{int}(t)$ in the system on and off during well-defined amounts of time by 
changing external parameters. This is often the case in 
real qubit systems where the qubits consist of massive particles such as electrons or ions. 
Our scheme is based on separating the two-qubit (entangling) part of $U$ from the single-qubit 
basis transformations (the so-called $K_1AK_2$ or single-value decomposition~\cite{krau01,khan01}, with $A\in SU(4)$ 
and $K_1, K_2 \in SU(2) \times SU(2)$), translating each of the matrices $K_1$, $A$, and $K_2$ into a  
sequence of operations which involves the smallest possible number of $U_{int}(t)$ for a suitably 
chosen time $t$, and minimizing the
number of single-qubit rotations in the total sequence of operations using 
permutation relations. The resulting shortest-sequence implementation of $U$ - which we refer in
this paper to as 
the {\it optimal} implementation of $U$ - contains the least number of operations from the 
given "library" of single- and two-qubit gates~\cite{vida02}. We illustrate our scheme by calculating the 
optimal implementation of three elementary quantum operations $U$ [the transformation
from standard basis to Bell basis, the CNOT gate, and the quantum Fourier transform] 
for two kinds of two-qubit interaction ${\cal H}_{int}$ that are relevant for solid-state qubits: 
Heisenberg exchange
interaction (corresponding to the (SWAP)$^{\al}$-gate) and quantum inductive coupling 
(corresponding to the B-gate). Either of these gates, in combination with single-qubit rotations, 
forms a so-called universal set~\cite{fan05,zhan04,burk99} into which any operator 
$U\in SU(4)$ can be decomposed. 

The paper is organized as follows. In Sec.~\ref{sec-protocol} we outline the decomposition protocol, 
starting with the decomposition of an arbitrary gate $U\in SU(4)$ into matrices 
$K_1$, $A$ and $K_2$. Sec.~\ref{subsec-max} addresses the special case in which $U$ is a maximally 
entangling gate, for which the protocol can be simplified. In Sec.~\ref{subsec-efficient} we show 
how each of the matrices $K_1$, $A$ and $K_2$ can be decomposed into a sequence of 
operations consisting of single-qubit rotations and a minimum number of 
two-qubit interactions $U_{int}(t)$ of a given kind.
In Sec.~\ref{subsec-optimization} we show how to iteratively minimize the total number of single-qubit rotations 
in the decomposition of $U$, and thereby obtain the implementation of $U$ that involves the 
smallest number of both $U_{int}(t)$ and single-qubit rotations. In section~\ref{sec-illustration}, 
we illustrate the decomposition protocol for two kinds of interaction that occur in 
solid-state qubit systems: Heisenberg exchange 
interaction ${\cal H}_{EX}$ and quantum inductive coupling ${\cal H}_{ind}$. We show how to 
implement the transformation from
standard basis to Bell basis (Sec.~\ref{subsec-stBell}), the CNOT-gate (Sec.~\ref{subsec-CNOT})
and the quantum Fourier transform (Sec.~\ref{subsec-F}) using either of these interactions. 
Finally, in Sec.~\ref{sec-application} we discuss how our results 
can be implemented for electron spin qubits in quantum dots and for superconducting flux qubits,
followed by conclusions in Sec.~\ref{sec-conclusion}.

\section{Explicit decomposition protocol}
\label{sec-protocol}

\subsection{General gates $U \in SU(4)$}
\label{subsec-gen}

The decomposition protocol we present in this section is based on the 
$K_1 A K_2$ decomposition~\cite{krau01,khan01}, which states that every two-qubit operation
$U \in SU(4)$ can be written as $U=K_1 A K_2$, with $K_1$, $K_2$ $\in SU(2) \otimes SU(2)$ and
\bea
A & = & \exp\left[ i(\al \sig_x \otimes \sig_x + \beta \sig_y \otimes \sig_y + \gam \sig_z \otimes \sig_z)\right]
\nn \\
& = & \left( \ba{cccc}
P_{\al \beta}^{-} e^{i \gamma} & 0 & 0 & i Q_{\al \beta}^{-} e^{i\gam} \\
0 &  P_{\al \beta}^{+} e^{-i \gamma} & i Q_{\al \beta}^{+} e^{-i \gam} & 0 \\
0 & i Q_{\al \beta}^{+} e^{-i \gamma} & P_{\al \beta}^{+} e^{-i \gam} & 0 \\
i Q_{\al \beta}^{-} e^{i\gam} & 0 & 0 & P_{\al \beta}^{-} e^{i \gam} 
\ea \right), 
\label{eq:Adef}
\eea
with $P_{\al \beta}^{\pm} \equiv \cos (\al \pm \beta)$ and $Q_{\al \beta}^{\pm} \equiv \sin (\al \pm \beta)$.
Here $\al, \beta, \gam \in [0, 2\pi)$ and $\sig_x$, $\sig_y$, $\sig_z$ denote the Pauli matrices. 
Our goals in this section are 1. to calculate the matrices $K_1$, $K_2$ and $A$ for a given $U \in SU(4)$ and 
2. to translate each of these matrices into the smallest number of single-qubit rotations and two-qubit 
interactions $U_{int}(t)$ of a given kind.
In this subsection and the next we develop a systematic protocol for the first goal~\cite{similar}. 
The implementation of
$K_1$, $A$ and $K_2$ using $U_{int}(t)$ and single-qubit rotations 
is the subject of section~\ref{subsec-efficient}.
\\
We begin by introducing the matrix $Q\in SU(4)$ defined as
\be
Q\equiv \frac{1}{\sqrt{2}} \left( 
\begin{array}{cccc}
1 & 0 & 0 & i \\ 
0 & i & 1 & 0 \\ 
0 & i & -1 & 0 \\ 
1 & 0 & 0 & -i
\end{array}
\right) \text{.}
\label{eq:Qdef}
\ee
$Q$ transforms the standard basis \{$|e_1\ra$, $|e_2\ra$, $|e_3\ra$, $|e_4\ra$\} of $SU(4)$
into the magic basis \{$|m_1\ra$, $|m_2\ra$, $|m_3\ra$, $|m_4\ra$\}. Here 
the standard basis is defined in spin-1/2 notation 
 with the spin along the $z$-direction as $|e_1\ra \equiv \left\vert \ua \ua\right\ra$, $|e_2\ra \equiv \left\vert \ua\da\right\ra$, 
$|e_3\ra \equiv \left\vert\da \ua \right\ra$, $|e_4\ra \equiv \left\vert\da \da\right\ra $
and the magic basis is defined as:
\begin{subequations}
\begin{eqnarray}
\left\vert m_{1}\right\rangle &\equiv &\frac{1}{\sqrt{2}}\left( \left\vert
\ua \ua\right\rangle +\left\vert \da \da\right\rangle \right) \\
\left\vert m_{2}\right\rangle &\equiv &\frac{1}{\sqrt{2}} i \left( \left\vert
\ua \da\right\rangle +\left\vert \da \ua\right\rangle \right) \\
\left\vert m_{3}\right\rangle &\equiv &\frac{1}{\sqrt{2}}\left( \left\vert
\ua \da\right\rangle -\left\vert \da \ua\right\rangle \right) \\
\left\vert m_{4}\right\rangle &\equiv &\frac{1}{\sqrt{2}}i\left( \left\vert
\ua \ua\right\rangle -\left\vert \da \da\right\rangle \right) \text{.}
\end{eqnarray}
\label{eq:magicbasis}
\end{subequations}
\\
Next, we define
\begin{eqnarray}
O_{1} &\equiv &Q^{\dag }K_{1}Q \label{eq:O1}\\
O_{2} &\equiv &Q^{\dag }K_{2}Q \label{eq:O2} \\
F &\equiv &Q^{\dag }AQ\text{,}
\label{eq:F}
\end{eqnarray}%
where $O_1$, $O_2$ $\in SO(4)$~\cite{SO4} and $F = {\rm diag}(\lam_1,\lam_2,\lam_3,\lam_4)$ with $\lam_1, \ldots \lam_4 \in 
\mathbb{C}$. Using (\ref{eq:Adef}), (\ref{eq:Qdef}) and (\ref{eq:F}), 
we find that the relation between $\lam_1, \ldots, \lam_4$ and
$\al, \beta, \gam$ is:
\begin{subequations}
\bea
\lam_1 &= & e^{i(\al - \beta + \gam)} \\
\lam_2 &= & e^{i(\al + \beta - \gam)} \\
\lam_3 &= & e^{i(-\al - \beta - \gam)} \\
\lam_4 &= & e^{i(-\al + \beta + \gam)}.
\eea 
\label{eq:lambdas}
\end{subequations}
Our aim~\cite{also} is to find the matrices $O_1$, $O_2$ and $F$, and from those via (\ref{eq:O1})-(\ref{eq:F}) the
matrices $K_1$, $K_2$ and $A$. To this end, we proceed in 3 steps: \\ \\
{\it Step 1:} We first calculate $U_{MB}$, where $U_{MB}$ is the representation 
of $U$ in the magic basis (\ref{eq:magicbasis}):
\be 
U_{MB} \equiv Q^{\dagger} U Q = O_1 F O_2.
\label{eq:UMB}
\ee 
We then define vectors $|v_i\ra$ as
\be
|v_i\ra \equiv O_2^{-1} |e_i\ra \ \ {\rm for}\ i=1 \ldots 4.
\label{eq:vi}
\ee
The vectors (\ref{eq:vi}) are also the eigenvectors of
$U_{MB}^{T} U_{MB}$ with corresponding
eigenvalues $\mu_i \equiv \lam_i^2$:
\bd
U_{MB}^{T}U_{MB}\left\vert v_{i}\right\rangle = O_{2}^{T}F^{2}O_{2}\left\vert
v_{i}\right\rangle =O_{2}^{T}F^{2}\left\vert e_{i}\right\rangle 
=  \lambda_{i}^{2}O_{2}^{T}\left\vert e_{i}\right\rangle =\mu _{i} \left\vert
v_{i}\right\rangle \text{.} 
\ed
Thus by calculating the eigenvalues and eigenvectors of $U_{MB}^{T} U_{MB}$, which is a known matrix,
we directly obtain the matrix $O_2$ from Eq.~(\ref{eq:vi}) as
\bd
O_2 = (|v_1\ra \ldots |v_4\ra)^{T}
\ed
and hence, using Eq.~(\ref{eq:O2}),
\be
K_2 = Q (|v_1\ra \ldots |v_4\ra)^{T} Q^{\dagger}.
\label{eq:K2expl}
\ee
In the above we are free to assign the eigenvectors of $U_{MB}^{T}U_{MB}$ 
in some chosen order to the vectors $\left\vert v_{i}\right\rangle$, 
with the restriction that since $O_{2}\in SO\left( 4\right)$ it should hold 
that $\det O_{2}=1$. \\ \\
{\it Step 2:} Using the eigenvalues $\lam_i^2$ of $U_{MB}^{T}U_{MB}$ found
in Step 1, we calculate the arguments $\al$, $\beta$ and $\gam$ of the
matrix $A$ [Eq.~(\ref{eq:Adef})]~\cite{others} using (\ref{eq:lambdas}) :
\begin{subequations}
\label{475}
\begin{eqnarray}
\alpha &=&\frac{-i}{4}\log \left( \lambda _{1}\lambda _{2}\lambda
_{3}^{-1}\lambda _{4}^{-1}\right) +\frac{1}{2}k_{1}\pi \\
\beta &=&\frac{-i}{4}\log \left( \lambda _{1}^{-1}\lambda _{2}\lambda
_{3}^{-1}\lambda _{4}\right) +\frac{1}{2}k_{2}\pi \\
\gamma &=&\frac{-i}{4}\log \left( \lambda _{1}\lambda _{2}^{-1}\lambda
_{3}^{-1}\lambda _{4}\right) +\frac{1}{2}k_{3}\pi \text{,}
\label{eq:gam}
\end{eqnarray}
\label{eq:albetagam}
\end{subequations}
where $k_{1},k_{2},k_{3}\in \mathbb{Z}$. \\ \\
{\it Step 3:} Finally, the matrix $O_{1}$ is calculated from
\bd
O_1 = O_{1}FO_{2}O_{2}^{T}F^{\ast }= U_{MB}O_{2}^{T}F^{\ast },
\ed
and hence
\begin{subequations}
\begin{eqnarray}
K_1 & = & Q U_{MB}O_{2}^{T}F^{\ast } Q^{\dagger} \\
& = & U (Q Q^{T}) K_2^{T} A^{*} (Q Q^{T})^{*}.
\end{eqnarray}
\label{eq:K1expl}
\end{subequations}
\noindent There are two degrees of freedom in this decomposition procedure. First, there are different ways 
of identifying the eigenvectors of $U_{MB}^{T}U_{MB}$ with the vectors $\left\vert v_{i}\right\rangle$. 
Secondly, there is some freedom in choosing the
arguments $\alpha $, $\beta $ and $\gamma $ of the operation $A$. In general, different
choices lead to different matrices $K_1$, $A$ and $K_2$, and the $K_1 A K_2$-decomposition of $U$
is not unique.

At this point it is useful to compare the procedure for obtaining a $K_1$$A$$K_2$-decomposition described 
above to the one proposed by Kraus and Cirac (KC)~\cite{krau01}. There are two main differences between
the two: 1) Using KC's method the 
matrix $K_2$ is given in the form SU(2) $\times$ SU(2) and involves 3 unknowns 
(Euler angles), and 2) KC's method leads to fixed values of the 
$\lambda_i$'s (instead of freedom of choice of the sign of each $\lambda_i$).
Which of the two methods is more convenient to use depends on 
the goal of the calculation to be made. If this goal is 
to (only) find a $K_1$$A$$K_2$ decomposition of a given matrix $U$, then 
having $K_2$ in SU(2) $\times$ SU(2) form and the $\lambda_i$'s fixed may 
be advantageous, if it leads to less equations to be solved. 
If, on the other hand, the goal of the calculation is to first 
find a $K_1$$A$$K_2$ decomposition of $U$ and then decompose each of 
the matrices $K_1$, $A$ and $K_2$ in terms of single-qubit rotations and a 
given tunable two-qubit interaction $U_{int}(t)$ (as it is in this paper), 
the freedom of assigning the vectors $|v_i\ra$ and choosing the sign 
of the $\lambda_i$'s that our method entails is advantageous, since 
it allows us to choose these parameters in such a way that the 
ensuing decomposition into single-qubit rotations and $U_{int}(t)$ 
is straightforward.

\subsection{Maximally entangling gates $U\in SU(4)$}
\label{subsec-max}

The procedure described in the previous subsection can be simplified if $U$ is a 
quantum gate which maximally entangles a suitably chosen separable basis, which we 
refer to as a "maximally entangling operator" or a "quantum gate with maximally entangling capacity"
(not to be confused with quantum gates which "only" maximally entangle a single separable 
quantum {\it state} instead of a whole {\it basis}). For this special class
of two-qubit operations the following Proposition holds:

\noi {\it Proposition 1}: Any transformation $M \in SU(4)$ from 
the standard basis to a maximally-entangled basis 
can be expressed in the following way:
\be
M = K_1\, e^{\frac{1}{4}i\pi (\sig_x \otimes \sig_x + \epsilon \sig_z \otimes \sig_z)}\, R_x^{(1)}(\mu) 
R_x^{(2)}(\nu) R_y^{(1)}(\xi)\, R_y^{(2)}(\zeta) R_z^{(1)}(\eta) R_z^{(2)}(\omega), 
\label{eq:U}
\ee
where $\eps$, $\mu$, $\nu$, $\xi$, $\zeta$, $\eta$, and $\omega$ satisfy one of the following three
conditions: 
\begin{subequations}
\bea
1) & \eps =0, &  \xi, \zeta \in \{0, \pm \pi\},\ \ \ \mu, \nu, \eta, \omega \in [0,2\pi)
\label{eq:eps0}  \\
2) & 0<\eps < 1, &  \mu, \nu, \xi, \zeta \in \{0, \pm \pi\},\ \  \eta, \omega \in [0,2\pi)
 \label{eq:eps01} \\
3) & \eps =1, & \mu, \nu \in \{0, \pm \pi\},\ \ \ \xi, \zeta, \eta, \omega \in [0,2\pi). 
\label{eq:eps1}
\eea
\label{eq:proposition}
\end{subequations}

\noi {\it Proof}: It has recently been shown~\cite{reza04} that any maximally entangling operator 
is of the form $A(\eps) \equiv \exp \left[ (1/4) i \pi (\sigma_{x}
\otimes \sigma _{x} + \eps \sigma _{z}\otimes \sigma _{z}) \right]$ (with
$\eps \in [0,1]$ and modulo permutations in $\sig_x$, $\sig_y$, $\sig_z$). 
We are thus left with the question which rotations may be performed on the standard
basis before applying $A(\eps)$, such that $A(\eps)$ acts as a maximally entangling operation. 
Since rotations around the 
$z$-axis only multiply each quantum state by a phasefactor, 
they are always allowed. It is then necessary and sufficient to investigate for which values of $\eps$, 
$\mu$, $\nu$, $\xi$ and $\zeta$ the transformation $T \equiv A(\eps) R_x^{(1)}(\mu) R_x^{(2)}(\nu) 
R_y^{(1)}(\xi) R_y^{(2)}(\zeta)$ corresponds to a 
maximally entangling operation. These values can be found in a straightforward way 
by determining for which values of $\eps$, $\mu$, $\nu$, $\xi$ and
$\zeta$ the concurrence~\cite{woot98} equals 1 for each 
of the four column vectors $|w_i\ra$. The calculation is given in the Appendix and 
leads to the solution Eq.~(\ref{eq:proposition}). 
$\blacksquare$ \\
Note that in order to transform from the standard basis to a {\it specific} maximally entangled basis 
the conditions on the angles in Eq.~(\ref{eq:proposition}) are in general more restrictive. For example, 
to transform to the Bell basis we require 
$\mu, \nu, \xi, \zeta$ $\in \{0, \pm \pi\}$, $\eta, \omega$ $\in [0,2\pi)$ and $\al_i, 
\beta_i, \gam_i$ $\in [0, \pm \pi/2, \pm \pi]$ for all values of $\eps$.
\\ \\
For gates $U\in SU(4)$ which maximally entangle the standard basis, step 2 in the decomposition procedure from 
the previous subsection can now be simplified: using Proposition 1 we choose $\lam_1$-$\lam_4$ 
such that $\beta=0$ and $\al=\pi/4$. For each of the possible assignments we then calculate 
$\gam$ (and thus the value of $\eps$ in Eq.~(\ref{eq:U})) from Eq.~(\ref{eq:gam}). 
\\
Having found the decomposition of $U$ into the matrices $K_1$, $K_2$ and $A$, we now proceed 
to develop a systematic scheme to calculate the implementation of each of these matrices using a 
sequence of single-qubit rotations (for $K_1$ and $K_2$) plus two-qubit interactions $U_{int}(t)$ 
of a given kind (for $A$). This is the topic of the next subsection.

\subsection{Decomposing $K_1$, $K_2$ and $A$ into $U_{int}(t)$ and single-qubit rotations}
\label{subsec-efficient}

We first consider gates $U\in SU(4)$ with maximally entangling capacity, as defined in 
the previous subsection.
\\ \\
{\it Decomposition of $K_2$ into single-qubit rotations. -} For maximally entangling two-qubit gates, 
a decomposition of the matrix $K_2$ into single-qubit rotations can be found in a straightforward
way by comparing the matrix $K_2$ obtained from (\ref{eq:K2expl}) to 
the sequence of rotations on the right-hand side of Eq.~(\ref{eq:U}), using the appropriate conditions
on the angles (\ref{eq:proposition}) that corresponds to the value of $\eps = 4\gamma/\pi$ 
calculated from Eq.~(\ref{eq:gam}). The resulting decomposition of $K_2$ is not unique, 
as can be seen from (\ref{eq:proposition}) where
different sequences of single-qubit rotations each correspond to a correct decomposition of $K_2$.
\\ \\
{\it Decomposition of $K_1$ into single-qubit rotations. -} $K_1$ can be translated into
single-qubit rotations using the Euler decomposition theorem~\cite{bare95}, which states that
every $K \in SU(2) \otimes SU(2)$ can be written as
\be 
K= \prod_{j=1,2} R_{z}^{(j)}(\al_j)\, R_{y}^{(j)}(\beta_j)\, R_{z}^{(j)}(\gam_j).
\label{eq:K}
\ee
Here $R_{z}^{(j)}(\alpha_j)$ represents a rotation of qubit $j$ over $\alpha_j$ around 
the $z$-axis and $\al_j$, $\beta_j$, 
$\gam_j$ $\in [0,2\pi)$. By substituting $K_1$ calculated from Eq.~(\ref{eq:K1expl}) into the
left-hand side of Eq.~(\ref{eq:K}) and solving the 
resulting set of equations, we directly obtain the values of the angles $\al_i$, $\beta_i$ and $\gam_i$. 
It is clear that, as for $K_2$, the decompositon of $K_1$ is not unique and that 
different sequences of rotations all yield a 
correct decomposition of $K_1$.
\\ \\
{\it Decomposition of $A$ into $U_{int}(t)$ and single-qubit rotations. -} Our next goal is to decompose the matrix $A$ 
[Eq.~(\ref{eq:Adef}) with the values of $\al$, $\beta$ and $\gam$ obtained from Eq.~(\ref{eq:albetagam})]
into a sequence of two-qubit interactions $U_{int}(t)$ and single-qubit rotations such that the
 smallest number of $U_{int}(t)$ is used. To this end, we first examine the entangling capacity of the given two-qubit 
gate $U_{int}(t)$ as a function of $t$, i.e. we investigate how many applications of $U_{int}(t)$ 
are needed to transform the standard basis into
a maximally entangled basis for a given $t$. For tunable two-qubit gates $U_{int}(t)$ there exists in general 
a value of $t$, say $t^{*}$, for which this minimum number of applications
is small. The value of $t^{*}$ is obtained by calculating a measure of bipartite entanglement, such as 
the concurrence~\cite{woot98}, for $U_{int}(t)$ applied to the standard basis vectors. In many cases $t^{*}$ can 
be determined 
directly from inspection of $U_{int}(t)$ (for $U_{int}(t)$ written in the standard basis). For example, 
if $U_{int}(t)$ is the (SWAP)$^{\al}$-gate (see Sec.~\ref{sec-illustration},
Eq.~(\ref{eq:exch})), $t^{*}$ is determined by $\al$. In this case two applications of (SWAP)$^{1/2}$ 
(plus a single-qubit rotation in between) are 
sufficient to implement $U$~\cite{zhang03_2,devi06}, since (SWAP)$^{1/2}$ maximally entangles two out of 
four standard basis states (and affects the other two basis states only trivially,
by multiplication with a phase factor). Therefore $t^{*} = \pi/(2\hbar J)$ (for constant $J$, see 
Eq.~(\ref{eq:alpha}) below).

Once  a sequence of operations, say $\tilde{A}$, that transforms the standard basis into a maximally 
entangled basis and contains the smallest possible number of $U_{int}$ (plus single-qubit rotations) 
has been found, $A$ and $\tilde{A}$ 
differ by local operations only. The latter can be calculated in the same way as for $K_1$,
by using the Euler decomposition~(\ref{eq:K}) of $A \tilde{A}^{-1}$. 

For two-qubit entangling gates $U\in SU(4)$ that do not have maximally entangling capacity, the decomposition
procedure of $K_1$, $A$ and $K_2$ into $U_{int}(t)$ and single-qubit rotations is the same as for
maximally entangling gates described above - except that 
the matrix $K_2$ now also has to be decomposed by using the Euler decomposition~(\ref{eq:K}), 
since Proposition 1 does not apply.

\subsection{Optimization}
\label{subsec-optimization}

By joining together the individual decompositions of $K_1$, $A$ and $K_2$ obtained via the method
described in the previous subsection, we have obtained a decomposition of $U=K_1 A K_2$ 
which consists of the {\it smallest} number of two-qubit 
interactions (plus a number of single-qubit rotations). This optimal decomposition is obtained 
by first optimizing the number of two-qubit gates and subsequently optimizing the number of 
single-qubit gates. The total number of rotations in this decomposition can 
often be reduced by using one or both of the following: 
\\ \\
(i) Commutation relations between $U_{int}(t)$ (or $\tilde{A}$, as defined in the previous subsection) 
and rotations,\\
(ii) Euler's theorem: writing rotations $R_{m_i}^{(j)}(\phi)$ as~\cite{possible} $R_{n_i}^{(j)}(\al_1)$ 
$R_{p_i}^{(j)}(\phi)$ $R_{n_i}^{(j)}(\al_2)$ with $n_i \neq p_i$, both $n_i$ and $p_i$
orthogonal to $m_i$, and $\al_{1}, \al_{2} \in \{\pm \pi/2 \}$ (and using (i) again). \\
Once both of these do not lead to a further reduction of
the number of single-qubit rotations, the minimum number of rotations has been found.\\

In practice, depending on experimental conditions, it may only be possible (or be easier) to implement 
rotations around certain axes and therefore be necessary to e.g. translate rotations around the $z$-axis into a 
sequence of rotations around axes in the ($x,y$)-plane. By implementing specific requirements such 
as these and using commutation relations to minimize the number of rotations we obtain the 
implementation of $U$ which requires the smallest number of operations from a given library 
of single- and two-qubit operations.

\section{Illustration of the decomposition protocol}
\label{sec-illustration}
 
In this section we illustrate the protocol developed in the previous section for two types
of interaction $U_{int}(t)$: Heisenberg exchange interaction and quantum inductive coupling. 
Heisenberg exchange interaction is described by the Hamiltonian
\bd
{\mathcal H}_{EX}(t) = (1/4) \hbar^2 J(t)\, \vec{\sig}^{(1)} \cdot \vec{\sig}^{(2)},
\ed
with $J(t)$ the time-dependent (tunable) exchange energy. This interaction corresponds
 to the (SWAP)$^{\al}$-gate: $({\rm SWAP})^{\al}$ $\equiv$ $U_{EX}(t)$ = 
$\exp [-(i/\hbar)\int_0^t {\mathcal H}_{EX}(\tau) d\tau]$ with
\be 
\al(t) \equiv -\frac{\hbar}{\pi} 
\int_0^t J(\tau) d\tau
\label{eq:alpha}
\ee
or, equivalently,
\be
({\rm SWAP})^{\al} = e^{-\frac{\al}{4}i\pi} \left(
\ba{cccc}
e^{\frac{\al}{2}i\pi} & 0 & 0 & 0 \\
0 & \cos (\frac{\al}{2}\pi) & i\sin (\frac{\al}{2}\pi) & 0 \\
0 & i \sin (\frac{\al}{2}\pi) & \cos (\frac{\al}{2}\pi) & 0 \\
0 & 0 & 0 & e^{\frac{\al}{2}i\pi}
\ea \right). 
\label{eq:exch}
\ee
The second type of interaction we use to illustrate our protocol is
a tunable coupling between two magnetic fluxes (see Section~\ref{sec-application} for a description
of a physical realization using so-called flux qubits). The Hamiltonian corresponding to
this type of interaction is given by~\cite{nisk06}
\be
{\cal H}_{\rm ind}(t) \equiv  - \frac{g_{+}(t)}{4} (\sigma_x \otimes
\sigma_x - \sig_y \otimes \sig_y) - \frac{g_{-}(t)}{4} (\sigma_x \otimes
\sigma_x + \sig_y \otimes \sig_y), 
\label{eq:Hamflux}
\ee
$\gam_{\pm} (t) \equiv \frac{1}{\hbar} \int_0^{t} g_{\pm}(\tau) 
d\tau$, 
where $g_{\pm}(t)$ are tunable system parameters. The Hamiltonian (\ref{eq:Hamflux}) 
corresponds to the so-called B-gate~\cite{zhan04}:
$B(\gam_{+},\gam_{-}) \equiv U_{\rm int} (t)
= \exp[-\frac{i}{\hbar} \int_0^t {\cal H}_{\rm int} (\tau)d\tau]$ or, equivalently,
\be
B(\gam_{+},\gam_{-}) =  \left(
\ba{cccc}
\cos (\frac{\gam_{+}}{2}) & 0 & 0 & i \sin (\frac{\gam_{+}}{2}) \\
0 & \cos (\frac{\gam_{-}}{2}) & i\sin (\frac{\gam_{-}}{2}) & 0 \\
0 & i \sin (\frac{\gam_{-}}{2}) & \cos (\frac{\gam_{-}}{2}) & 0 \\
 i \sin (\frac{\gam_{+}}{2}) & 0 & 0 & \cos (\frac{\gam_{+}}{2})
\ea \right) 
\label{eq:Bgate}
\ee 
with $\gam_{\pm} \in [0,2\pi)$. Note that Eq.~(\ref{eq:Bgate}) for $\gam_{+}, \gam_{-}=\pm 
\pi/2$ maximally entangles the 
entire standard basis, whereas the exchange interaction
$\sqrt{\rm SWAP}$ only produces maximal entanglement for two out of the four states in the standard basis. 
For both ${\cal H}_{EX}(t)$ and ${\cal H}_{\rm ind}(t)$ we now calculate the optimal (shortest-sequence) implementation 
of three elementary quantum operations: the transformation 
from standard basis to Bell basis, the CNOT gate, and the quantum Fourier transform. The former
two are maximally entangling operations, but the latter,
as shown below, is not.

\subsection{Transformation from standard to Bell basis}
\label{subsec-stBell}

A transformation from the standard basis to the Bell basis is 
by definition a maximally entangling operation. It has been shown that when using the
Heisenberg exchange
interaction [The (SWAP)$^{\al}$-gate, Eq.~(\ref{eq:exch})] the shortest 
sequence of operations that transforms the standard basis into 
a Bell basis is given by~\cite{burk99,devi06}
\bea
M_{n_{\phi}}^{(j)} & \equiv & \sqrt{\rm SWAP}\, R_{n_{\phi}}^{(j)}(\pi) 
\sqrt{\rm SWAP} \nn \\
& \stackrel{j=1}{=} & \frac{e^{-i\frac{\pi}{4}}}{\sqrt{2}}\, \left( \ba{cccc}
0 & e^{-i\phi} & -i e^{-i\phi} & 0 \\
e^{i\phi} & 0 & 0 & -i e^{-i\phi} \\
-i e^{i\phi} & 0 & 0 &  e^{-i\phi} \\
0 & -i e^{i\phi} & e^{i\phi} & 0 
\ea \right), 
\label{eq:UstBell}
\eea
where $j$=1,2 labels the qubit and $R_{n_{\phi}}^{(j)}(\pi)$ represents a rotation of qubit $j$ around 
an arbitrary axis $n_{\phi} \equiv (\cos \phi, \sin \phi, 0)$ in the $(x,y)$-plane. 
The analogous transformation for the quantum inductive coupling  
consists of a single application~\cite{makh02} of the $B$-gate [Eq.~(\ref{eq:Bgate})] with $\gam_{+}$, $\gam_{-} = 
\pm \frac{\pi}{2}$. In order to find the transformation from the standard basis to 
the "standard Bell basis" \{($1/\sqrt{2}) (\left\vert\ua \ua \ra \pm |\da \da\right\ra)$, 
($1/\sqrt{2}) (\left\vert\ua \da \ra \pm |\da \ua\right\ra)$\}, we need to decompose the matrix
\be
U^{\rm st\rightarrow Bell} = \frac{1}{\sqrt{2}}\left( 
\begin{array}{cccc}
1 & 0 & 0 & 1 \\ 
0 & 1 & 1 & 0 \\ 
0 & 1 & -1 & 0 \\ 
1 & 0 & 0 & -1
\end{array}
\right)  
\label{eq:sttoBell}
\ee
(or a permutation of (\ref{eq:sttoBell}) in which the columns are interchanged).
Using the protocol outlined in the previous section, we first calculate the matrices 
$K_1$, $A$, and $K_2$ and then decompose these matrices into the shortest sequence of 
$U_{int}$ = (SWAP)$^{\al}$ or $U_{int} = B$ plus single-qubit rotations.
\\ \\
Starting with Step 1 in Section~\ref{subsec-gen}, we find that
\bd
U_{MB}^{\rm st\rightarrow Bell} = \frac{1}{\sqrt{2}}\left( 
\begin{array}{cccc}
1 & 0 & 0 & i \\ 
0 & 1 & -i & 0 \\ 
0 & i & -1 & 0 \\ 
-i & 0 & 0 & -1
\end{array}
\right) 
\ed
and the eigenvalues and corresponding eigenvectors of ($U_{MB}^{\rm st\rightarrow Bell^T} 
U_{MB}^{\rm st\rightarrow Bell}$) are given by
\begin{equation}
\begin{array}{cc}
\mu_{1}=\mu_{2}=i, \ \ \left\vert a_{1}\right\rangle =\frac{1}{\sqrt{2}}
\left( 
\begin{array}{c}
1 \\ 
0 \\ 
0 \\ 
1
\end{array}
\right), & \left\vert a_{2}\right\rangle =\frac{1}{\sqrt{2}}
\left( 
\begin{array}{c}
0 \\ 
-1 \\ 
1 \\ 
0
\end{array}
\right) \\ 
&  \\ 
\mu_3 = \mu_4 = -i, \ \ \left\vert a_{3}\right\rangle =\frac{1}{\sqrt{2}}
\left( 
\begin{array}{c}
0 \\ 
1 \\ 
1 \\ 
0
\end{array}%
\right) & \left\vert a_{4}\right\rangle =\frac{1}{\sqrt{2}}
\left( 
\begin{array}{c}
1 \\ 
0 \\ 
0 \\ 
-1
\end{array}
\right).
\end{array}
\label{eq:eigenvalues}
\ee
We now choose  $|v_i\ra \equiv |a_i\ra$ $\forall i=1, \dots, 4$
and calculate the matrix $K_2$ using (\ref{eq:K2expl}):
\be
K_2^{\rm st\rightarrow Bell} = \frac{1}{\sqrt{2}}\left( 
\begin{array}{cccc}
0 & 0 & 0 & 1+i \\ 
0 & 0 & -1-i & 0 \\ 
0 & -1+i & 0 & 0 \\ 
1-i & 0 & 0 & 0
\end{array}
\right). 
\label{eq:K2sttoBell}
\ee
Next, we use the eigenvalues $\lam_i \equiv \sqrt{\mu_i}$, $i = 1,\dots, 4$ [Eq.~(\ref{eq:eigenvalues})]
to calculate the matrix $A$ from Eqns.~(\ref{eq:albetagam}) and (\ref{eq:Adef}). 
Choosing $\lam_1 = \sqrt{i}$, $\lam_2 = - \sqrt{i}$, $\lam_3 = \sqrt{-i}$, $\lam_4 = - \sqrt{-i}$ 
we find the solution $\al = \pi/4$, $\beta = \gam = 0$ and hence
\be
A^{\rm st\rightarrow Bell} =\frac{1}{\sqrt{2}}\left( 
\begin{array}{cccc}
1 & 0 & 0 & i \\ 
0 & 1 & i & 0 \\ 
0 & i & 1 & 0 \\ 
i & 0 & 0 & 1
\end{array}
\right).
\label{eq:AsttoBell}
\ee
Finally, we obtain $K_1$ from Eq.~(\ref{eq:K1expl}):
\be
K_1^{\rm st\rightarrow Bell} = \frac{1}{\sqrt{2}} \left( 
\begin{array}{cccc}
1-i & 0 & 0 & 0 \\ 
0 & -1+i & 0 & 0 \\ 
0 & 0 & -1-i & 0 \\ 
0 & 0 & 0 & 1+i
\end{array}
\right).  
\label{eq:K1sttoBell}
\ee
We now decompose each of the matrices $K_2^{\rm st\rightarrow Bell}$, $A^{\rm st\rightarrow Bell}$ 
and $K_1^{\rm st\rightarrow Bell}$ 
[Eqns.~(\ref{eq:K2sttoBell})-(\ref{eq:K1sttoBell})]
into a sequence of single-qubit rotations and (SWAP)$^{\al}$-gates, using the 
procedure described in Section~\ref{subsec-efficient}. Starting with $K_2^{\rm st\rightarrow Bell}$ and comparing 
the matrix (\ref{eq:K2sttoBell}) to the right-hand side of Eq.~(\ref{eq:U}) for condition (\ref{eq:eps0}), 
we find:
\be
K_2^{\rm st\rightarrow Bell} = R_{y}^{(1)}(\pi)\, R_{y}^{(2)}(\pi)\, R_z^{(1)}(\frac{\pi}{2}).
\label{eq:K2dec}
\ee
By comparing the matrices in Eqns.~(\ref{eq:AsttoBell}) and (\ref{eq:UstBell}), we directly obtain a decomposition 
of $A^{\rm st\rightarrow Bell}$, since for $n_{\phi}=x$ and $j=1$ the two matrices only differ by a spin flip of the first qubit (which exchanges the first and third row as well as the second and fourth row). Hence we find, disregarding a global phasefactor,
\be
A^{\rm st\rightarrow Bell}  = R_{x}^{(1)}(\pi)\,  \sqrt{\rm SWAP}\, R_{x}^{(1)}(\pi)\, \sqrt{\rm SWAP}.
\label{eq:Adec}
\ee
This decomposition of $A^{\rm st\rightarrow Bell}$ is also obtained by following the decomposition procedure described in Sec.~\ref{subsec-efficient}, by first calculating the entangling capacity of (SWAP)$^{\al}$ $\forall \al$. 
Finally, the decomposition of the matrix $K_1^{\rm st\rightarrow Bell} \in SU(2) \otimes SU(2)$ is obtained 
by using Eq.~(\ref{eq:K}), and we find:
\be
K_1^{\rm st\rightarrow Bell} = R_{z}^{(1)}(-\frac{\pi}{2})\, R_z^{(2)}(\pi).
\label{eq:K1dec}
\ee
The total decomposition of $U^{\rm st\rightarrow Bell}$ in terms of $\sqrt{\rm SWAP}$-operations 
then becomes:
\bea
U^{\rm st\rightarrow Bell} & = &  K_1^{\rm st\rightarrow Bell} A^{\rm st\rightarrow Bell} 
K_2^{\rm st\rightarrow Bell}
\nn \\
& = & R_{z}^{(1)}(-\frac{\pi}{2})\, R_z^{(2)}(\pi)\,  
R_{x}^{(1)}(\pi)\,  \sqrt{\rm SWAP}\, R_{x}^{(1)}(\pi)\, \sqrt{\rm SWAP}\, 
R_{y}^{(1)}(\pi)\, R_{y}^{(2)}(\pi)\, R_z^{(1)}(\frac{\pi}{2}) \nn \\
& = & R_{z}^{(1)}(-\frac{\pi}{2})\,   
\sqrt{\rm SWAP}\, R_{x}^{(1)}(\pi)\, \sqrt{\rm SWAP}\, R_{y}^{(1)}(\pi)\, R_z^{(1)}(\frac{\pi}{2}) 
\label{eq:stBellfull}
\eea
In the last step of Eq.~(\ref{eq:stBellfull}) we have used the relation
$\sqrt{\rm SWAP}\, R_{x}^{(1)}(\pi)\, \sqrt{\rm SWAP}\, R_{y}^{(2)}(\pi)$ = 
$ R_z^{(2)}(\pi)\, R_{x}^{(1)}(\pi)\,  \sqrt{\rm SWAP}\, R_{x}^{(1)}(\pi)\, 
\sqrt{\rm SWAP}$. Eq.~(\ref{eq:stBellfull}) is the shortest sequence of operations that can be used 
to implement $U^{\rm st\rightarrow Bell}$
using Heisenberg exchange interaction and single-qubit rotations. Repeating the
same procedure for other transformations from the standard basis to the Bell basis 
(obtained by permutations of the columns of Eq.~(\ref{eq:sttoBell})) we find the
following possible optimal decompositions of $U^{\rm st\rightarrow Bell}$: 
\be
U^{\rm st \rightarrow Bell}_{\rm (SWAP)^{\al}}  =  \left\{ \ba{l}
R_z^{(1)}(\sig \frac{\pi}{2}) M_x^{(i)} R_z^{(1)}(\pm \frac{\pi}{2}) \\
R_z^{(1)}(-\sig \frac{\pi}{2}) R_x^{(1)}(\pi) M_x^{(i)} R_z^{(1)}(\pm \frac{\pi}{2}) \\ 
R_z^{(1)}(\sig \frac{\pi}{2}) R_y^{(1)}(\pi) M_x^{(i)} R_z^{(1)}(\pm \frac{\pi}{2}) \\ 
R_z^{(1)}(-\sig \frac{\pi}{2}) M_x^{(i)} R_{x,y}^{(1)}(\pi) R_z^{(1)}(\pm \frac{\pi}{2})
\ea \right.
\label{eq:UstBellspin}
\ee
with $(i,\sig)$ $\in$ $\{(1,1),(2,-1)\}$~\cite{zrotations}. 
\\
The analogue of Eq.~(\ref{eq:UstBellspin}) for quantum induced coupling (the B-gate,
Eq.~(\ref{eq:Bgate})) is given by:
\be
U^{\rm st \rightarrow Bell}_{\rm B}  =  \left\{ \ba{l}
R_z^{(1)}(-\sig \frac{\pi}{2}) B(\sig \frac{\pi}{2},\sig \frac{\pi}{2})  R_z^{(1)}(\pm \frac{\pi}{2}) \\
R_z^{(1)}(\sig \frac{\pi}{2}) R_x^{(1)}(\pi) B(\sig \frac{\pi}{2},\sig \frac{\pi}{2})  R_z^{(1)}(\pm \frac{\pi}{2}) \\
R_z^{(1)}(-\sig \frac{\pi}{2}) R_x^{(1)}(\pi) B(\sig \frac{\pi}{2},\sig \frac{\pi}{2}) 
R_{x,y}^{(1)}(\pi) R_z^{(1)}(\pm \frac{\pi}{2}) \\
R_z^{(1)}(\sig \frac{\pi}{2}) R_y^{(1)}(\pi) B(\sig \frac{\pi}{2},\sig \frac{\pi}{2}) 
R_{x}^{(1)}(\pi) R_z^{(1)}(\pm \frac{\pi}{2}) \\
R_z^{(1)}(\sig \frac{\pi}{2}) B(\sig \frac{\pi}{2},\sig \frac{\pi}{2}) 
R_{x}^{(1)}(\pi) R_z^{(1)}(\pm \frac{\pi}{2})
\ea \right. 
\label{eq:UstBellflux}
\ee
with $\sig = \pm 1$. Note that Eqns.~(\ref{eq:UstBellspin}) and (\ref{eq:UstBellflux}) involve six 
rotations more than the decomposition of $M_{n_{\phi}}$, Eq.~(\ref{eq:UstBell}). From a practical point of view, 
it is thus more efficient to implement a transformation from the standard basis to a Bell basis 
with complex coefficients 
than to the ``standard Bell basis" when either $U_{int}=$(SWAP)$^{\alpha}$ 
or $U_{int}=B$ is used.

To conclude this section, we also give the decomposition of two alternative matrices $A^{\rm st \rightarrow Bell}$.
The first one is obtained by using the $K_1 A K_2$- decomposition procedure of Kraus and Cirac~\cite{krau01}
and reads:
\be
A^{\rm st \rightarrow Bell}_{\rm alt 1} = 
\left( 
\begin{array}{cccc}
0 & 0 & 0 & 1 \\ 
0 & 0 & -i & 0 \\ 
0 & -i & 0 & 0 \\ 
1 & 0 & 0 & 0
\end{array}
\right).
\label{eq:alt1}
\ee
$A^{\rm st \rightarrow Bell}_{\rm alt 1}$ does not maximally entangle the standard basis. In order to find its 
decomposition, we first calculate an unentangled basis for which $A^{\rm st \rightarrow Bell}_{\rm alt 1}$ does
act as a maximal entangler. To this end, we note from the matrix (\ref{eq:alt1}) that the latter basis must be obtained 
from the standard basis by creating a superposition of both the first and the second qubit, since Eq.~(\ref{eq:alt1})
couples $|\ua \ua\ra$ to $|\da \da\ra$ and $|\ua \da\ra$ to $|\da \ua \ra$. Mathematically, a possible choice of 
rotations is 
\bd
\left( 
\begin{array}{cccc}
1 & -1 & -1 & 1 \\ 
1 & 1 & -1 & -1 \\ 
-1 & 1 & -1 & 1 \\ 
-1 & -1 & -1 & -1
\end{array}
\right),
\ed
which translates into (using Euler's decomposition theorem) $R_y^{(1)}(\frac{\pi}{2})\, R_z^{(1)}(\pi)\, R_y^{(2)}(-\frac{\pi}{2})$.
Thus $A^{\rm st \rightarrow Bell}_{\rm alt 1}$ $R_y^{(1)}(\frac{\pi}{2})\, R_z^{(1)}(\pi)\, R_y^{(2)}(-\frac{\pi}{2})$ =
(basis rotation) $\times$ $\sqrt{\rm SWAP}\, R_{x}^{(1)}(\pi)\, \sqrt{\rm SWAP} \equiv$ (basis rotation) $\times$ $M_x^{(1)}$.
The remaining basis rotation is easily found by using Euler's theorem and the full decomposition then reads
\bd
A^{\rm st \rightarrow Bell}_{\rm alt 1} = R_y^{(1)}(\frac{\pi}{2})\, R_y^{(2)}(\frac{\pi}{2})\, M_x^{(1)}\,
R_z^{(1)}(\pi)\, R_y^{(1)}(-\frac{\pi}{2})\, R_y^{(2)}(\frac{\pi}{2}).
\ed

As a second example of an alternative matrix $A^{\rm st \rightarrow Bell}_{\rm alt 2}$, we consider the matrix
obtained by using a different choice of $\lam_{i}$'s, namely $\lam_1=\sqrt{-i}$, $\lam_2=-\sqrt{i}$, $\lam_3=\sqrt{i}$
and $\lam_4=-\sqrt{-i}$. For this choice the matrix $A$ [Eq.~(\ref{eq:Adef})] becomes
\be
A^{\rm st \rightarrow Bell}_{\rm alt 2} = 
\left( 
\begin{array}{cccc}
1 & 0 & 0 & 0 \\ 
0 & -i & 0 & 0 \\ 
0 & 0 & -i & 0 \\ 
0 & 0 & 0 & 1
\end{array}
\right).
\label{eq:alt2}
\ee
$A^{\rm st \rightarrow Bell}_{\rm alt 2}$ acts as a maximal entangler on the same unentangled basis as
$A^{\rm st \rightarrow Bell}_{\rm alt 1}$ (see above), so that $A^{\rm st \rightarrow Bell}_{\rm alt 2}$ 
$R_y^{(1)}(\frac{\pi}{2})\, R_z^{(1)}(\pi)\, R_y^{(2)}(-\frac{\pi}{2})$ =
(basis rotation) $\times$ $M_x^{(1)}$. The remaining basis rotation is again found by decomposing 
$A^{\rm st \rightarrow Bell}_{\rm alt 2}$ $R_y^{(1)}(\frac{\pi}{2})\, R_z^{(1)}(\pi)\, R_y^{(2)}(-\frac{\pi}{2})$
$(M_x^{(1)})^{-1}$ using Euler's theorem and we find 
$A^{\rm st \rightarrow Bell}_{\rm alt 2}$ = $R_y^{(1)}(-\frac{\pi}{2})$ $R_y^{(2)}(-\frac{\pi}{2})$ $M_x^{(1)}\,$
$R_z^{(1)}(\pi)$ $R_y^{(1)}(-\frac{\pi}{2})$ $R_y^{(2)}(\frac{\pi}{2})$.

\subsection{The CNOT gate}
\label{subsec-CNOT} 

The CNOT-gate is given by
\bd
{\rm CNOT}^{(1,2)} = \left( 
\begin{array}{cccc}
1 & 0 & 0 & 0 \\ 
0 & 1 & 0 & 0 \\ 
0 & 0 & 0 & 1 \\ 
0 & 0 & 1 & 0
\end{array}
\right)
\ed
In order to find the optimal (shortest-sequence) decomposition of the CNOT gate in terms 
of the (SWAP)$^{\al}$ and B-gate, we again use the protocol from Sec.~\ref{sec-protocol} 
and proceed in the same way as for $U^{\rm st \rightarrow Bell}$ 
in the previous subsection. We first calculate $K_1$, $A$ and $K_2$, then decompose each of 
these matrices into (SWAP)$^{\al}$ or B-operations plus single-qubit rotations, and subsequently 
use permutation relations to optimize the number of rotations.
\\ \\
Starting with Step 1 in Section~\ref{subsec-gen}, we find for the representation of
${\rm CNOT}^{(1,2)}$ in the magic basis:
\bd
{\rm CNOT}^{(1,2)}_{MB} = \frac{1}{2}\left( 
\begin{array}{cccc}
1 & i & -1 & i \\ 
-i & 1 & -i & -1 \\ 
-1 & i & 1 & i \\ 
-i & -1 & -i & 1
\end{array}
\right) 
\ed
and the eigenvalues and corresponding eigenvectors of (${\rm CNOT}^{(1,2)^T}_{MB}
{\rm CNOT}^{(1,2)}_{MB}$) are given by
\begin{equation}
\begin{array}{cc}
\mu_{1}=\mu_{2}=1, \ \ \left\vert a_{1}\right\rangle =\frac{1}{\sqrt{2}}
\left( 
\begin{array}{c}
0 \\ 
-1 \\ 
0 \\ 
1
\end{array}
\right), & \left\vert a_{2}\right\rangle =\frac{1}{\sqrt{2}}
\left( 
\begin{array}{c}
-1 \\ 
0 \\ 
1 \\ 
0
\end{array}
\right) \\ 
&  \\ 
\mu_3 = \mu_4 = -1, \ \ \left\vert a_{3}\right\rangle =\frac{1}{\sqrt{2}}
\left( 
\begin{array}{c}
1 \\ 
0 \\ 
1 \\ 
0
\end{array}
\right) & \left\vert a_{4}\right\rangle =\frac{1}{\sqrt{2}}
\left( 
\begin{array}{c}
0 \\ 
1 \\ 
0 \\ 
1
\end{array}
\right).
\end{array}
\label{eq:eigenvaluesCNOT}
\ee
We now choose $|v_1\ra \equiv |a_3\ra$, $|v_2\ra \equiv |a_4\ra$,
$|v_3\ra \equiv |a_2\ra$ and $|v_4\ra \equiv |a_1\ra$,
and calculate the matrix $K_2$ from Eq.~(\ref{eq:K2expl}):
\be
K_2^{\rm CNOT} = \frac{1}{\sqrt{2}}\left( 
\begin{array}{cccc}
1 & 0 & -1 & 0 \\ 
0 & 1 & 0 & -1 \\ 
1 & 0 & 1 & 0 \\ 
0 & 1 & 0 & 1
\end{array}
\right). 
\label{eq:K2CNOT}
\ee
Next, we calculate the matrix $A$ from Eqns.~(\ref{eq:albetagam}) and (\ref{eq:Adef}). 
Choosing $\lam_1 = \lam_2 = 1$, $\lam_3 = \lam_4 = i$,
we find the solution $\al = \pi/4$, $\beta = \gam = 0$ and hence
\be
A^{\rm CNOT} =\frac{1}{\sqrt{2}}\left( 
\begin{array}{cccc}
1 & 0 & 0 & i \\ 
0 & 1 & i & 0 \\ 
0 & i & 1 & 0 \\ 
i & 0 & 0 & 1
\end{array}
\right).
\label{eq:ACNOT}
\ee
Finally, we obtain $K_1$ from Eq.~(\ref{eq:K1expl}):
\be
K_1^{\rm CNOT} = \frac{1}{2} \left( 
\begin{array}{cccc}
1 & -i & 1 & -i \\ 
-i & 1 & -i & 1 \\ 
-i & -1 & i & 1 \\ 
-1 & -i & 1 & i
\end{array}
\right).  
\label{eq:K1CNOT}
\ee
We now decompose each of the matrices $K_2^{\rm CNOT}$, $A^{\rm CNOT}$ 
and $K_1^{\rm CNOT}$ 
[Eqns.~(\ref{eq:K2CNOT})-(\ref{eq:K1CNOT})]
into a sequence of single-qubit rotations and (SWAP)$^{\al}$-gates, using the 
procedure described in Section~\ref{subsec-efficient}. Starting with $K_2^{\rm CNOT}$ and comparing 
the right-hand sides of Eqns.~(\ref{eq:K2CNOT}) and (\ref{eq:U}) for condition (\ref{eq:eps0}), 
we obtain:
\be
K_2^{\rm CNOT} = R_{y}^{(1)}(\frac{\pi}{2}).
\label{eq:K2CNOTdec}
\ee
Since $A^{\rm CNOT} = A ^{\rm st \rightarrow Bell}$, the decompositions of these matrices 
are the same:
\be
A^{\rm CNOT}  = R_{x}^{(1)}(\pi)\,  \sqrt{\rm SWAP}\, R_{x}^{(1)}(\pi)\, \sqrt{\rm SWAP}.
\label{eq:ACNOTdec}
\ee
Finally, the decomposition of the matrix $K_1^{\rm CNOT}$ $\in$ $SU(2) \otimes SU(2)$ is obtained 
using Eq.~(\ref{eq:K}), and we find:
\be
K_1^{\rm CNOT} = R_{y}^{(1)}(-\frac{\pi}{2})\, R_x^{(1)}(\frac{\pi}{2})\, R_x^{(2)}(\frac{\pi}{2}).
\label{eq:K1CNOTdec}
\ee
The total decomposition of ${\rm CNOT}^{(1,2)}$ is then given by:
\bea
{\rm CNOT}_{\rm (SWAP)}^{(1,2)} & = &  K_1^{\rm CNOT} A^{\rm CNOT} 
K_2^{\rm CNOT}
\nn \\
& = & R_{y}^{(1)}(-\frac{\pi}{2})\, R_x^{(2)}(\frac{\pi}{2})\,  
R_{x}^{(1)}(-\frac{\pi}{2})\,  \sqrt{\rm SWAP}\, R_{x}^{(1)}(\pi)\, \sqrt{\rm SWAP}\, R_{y}^{(1)}(\frac{\pi}{2}). 
\label{eq:CNOTfull}
\eea
The number of operations in Eq.~(\ref{eq:CNOTfull}) cannot be further reduced by applying commutation relations or Euler's theorem. Including possible permutations of rotation angles, we obtain the general form
of Eq.~(\ref{eq:CNOTfull})~\cite{zdecomp}:
\be
{\rm CNOT}_{\rm (SWAP)^{\al}}^{(1,2)} = 
R_y^{(1)}(\sigma \frac{\pi}{2})  R_x^{(i)}(\sigma^{\prime} \frac{\pi}{2})\,
R_x^{(j)}(\sig^{\prime \prime} \frac{\pi}{2}) 
\sqrt{\rm SWAP}\, R_x^{(i)}(\pi)\, \sqrt{\rm SWAP}\, R_y^{(1)}(- \sigma \frac{\pi}{2})   
\label{eq:CNOTspin}
\ee
for $(i,j,\sigma^{\prime},\sig^{\prime \prime})$ $\in$ $\{ (1,2,-1,-\sig), (2,1,\sig,1)\}$, with 
$\sig = \pm 1$. 
Using the B-gate, the shortest-sequence implementation of the CNOT$^{(1,2)}$-gate is given by:
\be
{\rm CNOT}_{\rm B}^{(1,2)} = 
R_y^{(1)}(\sigma \frac{\pi}{2})  R_x^{(1)}(\sigma^{\prime} \frac{\pi}{2}) 
R_x^{(2)}(-\sig^{\prime} \sig  \frac{\pi}{2})\, B(\sig^{\prime} \frac{\pi}{2},
\sig^{\prime} \frac{\pi}{2}) R_y^{(1)}(- \sigma \frac{\pi}{2}),
\label{eq:CNOTflux}
\ee
for $\sig, \sig^{\prime} = \pm 1$. The sequences Eqns.~(\ref{eq:UstBellspin}), (\ref{eq:UstBellflux}), (\ref{eq:CNOTspin}) and (\ref{eq:CNOTflux}) 
for implementing $U^{\rm st \rightarrow Bell}$ and the CNOT-gate are building blocks for the
implementation of quantum operations for three or more qubits~\cite{bodo07}.

\subsection{The quantum Fourier transform}
\label{subsec-F}

The quantum Fourier transform (QFT) lies at the heart of Shor's factoring algorithm~\cite{shor97} 
and is given by (for two qubits)~\cite{niel00}:
\be
{\cal F} = \frac{1}{2} \left( \ba{cccc}
1 & 1 & 1 & 1   \\
1 & i & -1 & -i \\            
1 & -1 & 1 & -1 \\
1 & -i& -1 & i 
\ea
\right).
\label{eq:QFT}
\ee
So far, the QFT has been implemented in an NMR system~\cite{vand00}, using ion 
qubits~\cite{chia05} and using phononic qubits~\cite{lu07}, but not yet with solid-state qubits. 
${\cal F}^2 
=$ CNOT$^{(2,1)}$ is a maximally entangling gate, but ${\cal F}$ itself is not~\cite{maxconc}.
In order to find the shortest sequence of operations required to implement ${\cal F}$,
we thus cannot use our optimization protocol for gates with maximally entangling capacity 
[as defined and described in Section~\ref{subsec-max}], but need to use the method for general gates $U\in SU(4)$
outlined in Section~\ref{subsec-gen}. To this end, we use the decomposition of ${\cal F}$
into the controlled phase gate $CP$ plus two Hadamard gates proposed by Coppersmith 
{\it et al.}~\cite{copp94}:
\be
{\cal F} = {\rm SWAP}\, \cdot H^{(2)} \cdot CP \cdot H^{(1)}
\label{eq:FCP}
\ee
with
\bse
\bea
H^{(2)} & = & \frac{1}{\sqrt{2}} \left( \ba{cccc}
1 & 1 & 0 & 0 \\
1 & -1 & 0 & 0 \\
0 & 0 & 1 & 1 \\
0 & 0 & 1 & -1 
\ea \right) \\
CP & = & \left( \ba{cccc}
1 & 0 & 0 & 0 \\
0 & 1 & 0 & 0 \\
0 & 0 & 1 & 0 \\
0 & 0 & 0 & i 
\ea \right) \label{eq:CP} \\
H^{(1)} & = & \frac{1}{\sqrt{2}} \left( \ba{cccc}
1 & 0 & 1 & 0 \\
0 & 1 & 0 & 1 \\
1 & 0 & -1 & 0 \\
0 & 1 & 0 & -1 
\ea \right).
\eea
\ese
The Hadamard gates $H^{(1)}$ and $H^{(2)}$ can be decomposed into
single-qubit rotations using Eq.~(\ref{eq:K}) and we find
\bea
H^{(i)} & = & R_x^{(i)} (\pi) R_y^{(i)}(\frac{\pi}{2}) \nn \\
& = & R_y^{(i)} (-\frac{\pi}{2}) R_x^{(i)}(\pi), \hspace*{0.5cm} i=1,2.
\label{eq:Hadamarddecomp}
\eea
We now proceed to calculate the decomposition $K_1 A K_2$ of $CP$ 
and to translate each of these matrices into single-qubit
rotations and two-qubit interactions according to the protocol given in
Sec.~\ref{sec-protocol}. Starting with $CP_{MB}$, the representation of $CP$ 
in the magic basis, we obtain from Eqns.~(\ref{eq:CP}), (\ref{eq:UMB}) and
(\ref{eq:Qdef}):
\bd
CP_{MB} = \frac{1}{2} \left( \ba{cccc}
1+i & 0 & 0 & 1+i   \\
0 & 2 & 0 & 0 \\            
0 & 0 & 2 & 0 \\
-1-i & 0 & 0 & 1+i 
\ea
\right).
\ed
The eigenvalues and eigenvectors of $CP_{MB}^{T} CP_{MB}$ are given by
\begin{equation}
\begin{array}{cc}
\mu_{1}=\mu_{2}= i, \ \ \left\vert a_{1}\right\rangle =
\left( 
\begin{array}{c}
1 \\ 
0 \\ 
0 \\ 
0
\end{array}
\right), & \left\vert a_{2}\right\rangle =
\left( 
\begin{array}{c}
0 \\ 
0 \\ 
0 \\ 
1
\end{array}
\right) \\ 
&  \\ 
\mu_3 = \mu_4 = 1, \ \ \left\vert a_{3}\right\rangle =
\left( 
\begin{array}{c}
0 \\ 
1 \\ 
0 \\ 
0
\end{array}
\right) & \left\vert a_{4}\right\rangle =
\left( 
\begin{array}{c}
0 \\ 
0 \\ 
1 \\ 
0
\end{array}
\right).
\end{array}
\label{eq:eigenvaluesF}
\ee
By choosing $|v_i\ra \equiv |a_i\ra$ for $i=1,\ldots 4$
and using Eq.~(\ref{eq:K2expl}), we obtain the matrix
$K_2^{CP}$:
\bea
K_2^{CP} & = & \frac{1}{2}\left( 
\begin{array}{cccc}
1 & i & -i & 1 \\ 
1 & -i & -i & -1 \\ 
1 & i & i & -1 \\ 
1 & -i & i & 1
\end{array}
\right) \nn \\
& = &  R_y^{(1)}(\frac{\pi}{2})\, 
 R_y^{(2)}(\frac{\pi}{2})\, R_z^{(1)}(\frac{\pi}{2})\, R_z^{(2)}(-\frac{\pi}{2})
 \nn \\
& = &  R_x^{(1)}(\frac{\pi}{2})\, 
 R_y^{(1)}(\frac{\pi}{2})\, R_x^{(2)}(-\frac{\pi}{2})\, R_y^{(2)}(\frac{\pi}{2}),
\label{eq:K2F}
\eea
where we have used Eq.~(\ref{eq:K}) to obtain the decomposition of $K_2^{CP}$.
Assigning $\lam_1 = \lam_2 = \sqrt{i}$ and
$\lam_3 = \lam_4 = 1$, we obtain from Eq.~(\ref{eq:albetagam}) the 
solution $\al = \pi/8$ and $\beta=\gam=0$, and thus from Eq.~(\ref{eq:Adef}):
\bea
A^{CP} & = & \left( 
\begin{array}{cccc}
\cos (\frac{\pi}{8}) & 0 & 0 & i\sin(\frac{\pi}{8}) \\ 
0 & \cos (\frac{\pi}{8})  & i\sin(\frac{\pi}{8})  & 0 \\ 
0 & i\sin(\frac{\pi}{8})  & \cos (\frac{\pi}{8})  & 0 \\ 
i\sin(\frac{\pi}{8})  & 0 & 0 & \cos (\frac{\pi}{8}) 
\end{array}
\right) \label{eq:AF1} \\
& = & R_{x}^{(1)}(\pi)\, (\rm SWAP)^{\frac{1}{4}}\, R_{x}^{(1)}(\pi)\, (\rm SWAP)^{\frac{1}{4}}.
\label{eq:AF2}
\eea
The decomposition (\ref{eq:AF2}) of $A^{CP}$ is obtained by calculating the concurrence (see
Eq.~(\ref{eq:concurrence})) of the column vectors of $A^{CP}$, which yields $C=4 \cos^2(\pi/8)
\sin^2(\pi/8) = 1/2$. Following the method outlined in Sec.~\ref{subsec-efficient}, we then calculate
the corresponding time $t^{*}$ such that (SWAP)$^{\al(t^{*})}$ yields the same value 1/2 of the concurrence 
(amount of entanglement) when applied to the standard basis vectors. We find that two applications of the (SWAP)$^{\frac{1}{4}}$-interaction, each followed by
a rotation of the first qubit over $\pi$ around the $x$-axis (which interchanges
the first and the third row, and the second and the fourth row of the matrix to which
it is applied) yield a decomposition of the matrix (\ref{eq:AF1}). \\
Having found the matrix $A^{CP}$, the matrix $K_1^{CP}$ and its decomposition 
can be calculated from Eqns.~(\ref{eq:K1expl}) and (\ref{eq:K}), respectively,
\bea
K_1^{CP} & = & \frac{1}{2} \left( 
\begin{array}{cccc}
1 & 1 & 1 & 1 \\
-i e^{\frac{1}{4} i \pi} & i e^{\frac{1}{4} i \pi} & -i e^{\frac{1}{4} i \pi} &
i e^{\frac{1}{4} i \pi} \\
i e^{\frac{1}{4} i \pi} & i e^{\frac{1}{4} i \pi} & -i e^{\frac{1}{4} i \pi} &
-i e^{\frac{1}{4} i \pi} \\
i & -i & -i & i  
\end{array}
\right) 
\nn \\
& = & R_z^{(2)}(\frac{3\pi}{4})\, 
 R_y^{(2)}(-\frac{\pi}{2})\, R_z^{(1)}(-\frac{\pi}{4})\, R_y^{(1)}(-\frac{\pi}{2})
 \nn \\
& = & R_y^{(1)}(-\frac{\pi}{2})\, 
 R_x^{(1)}(-\frac{\pi}{4})\, R_y^{(2)}(-\frac{\pi}{2})\, R_x^{(2)}(\frac{3\pi}{4}).
\label{eq:K1F2}
\eea
The decomposition of the quantum Fourier transform ${\cal F}$ is
then given by:
\bea
{\cal F} & = & {\rm SWAP}\ R_x^{(2)}(-\frac{\pi}{4})\, R_y^{(1)}(-\frac{\pi}{2})\, R_x^{(1)}(\frac{3\pi}{4})\,
({\rm SWAP})^{\frac{1}{4}}\, R_x^{(1)}(\pi)\, ({\rm SWAP})^{\frac{1}{4}}\, R_x^{(1)}(-\frac{\pi}{2})\, R_x^{(2)}(-\frac{\pi}{2})\, R_y^{(2)}(\frac{\pi}{2})
\nn \\
& = & R_x^{(1)}(-\frac{\pi}{4})\, R_y^{(2)}(-\frac{\pi}{2})\, R_x^{(2)}(\frac{3\pi}{4})\,
({\rm SWAP})^{\frac{5}{4}}\, R_x^{(1)}(\pi)\, ({\rm SWAP})^{\frac{1}{4}}\, R_x^{(1)}(-\frac{\pi}{2})\, R_x^{(2)}(-\frac{\pi}{2})\, R_y^{(2)}(\frac{\pi}{2}).
\label{eq:Fdecomp}
\eea
Using commutation relations to minimize the number of rotations in Eq.~(\ref{eq:Fdecomp}), we obtain the decomposition
of ${\cal F}$ that contains the smallest number of (SWAP)$^{\al}$- and single-qubit gates:
\be
{\cal F}_{\rm (SWAP)^{\al}} = R_y^{(2)}(-\frac{\pi}{2})
 R_x^{(2)}(\frac{\pi}{4})\, ({\rm SWAP})^{\frac{5}{4}}\,
R_x^{(1)}(\pi)\, ({\rm SWAP})^{\frac{1}{4}}\, R_x^{(2)}(-\frac{3\pi}{4}) R_y^{(2)}(\frac{\pi}{2}),  
\label{eq:QFTspin}
\ee
and analogously we find for the optimal decomposition of ${\cal F}$ using the B-gate~\cite{swapQFT}:
\be
{\cal F}_{B} = R_y^{(2)}(-\frac{\pi}{2})\, B(\frac{\pi}{4},\frac{5\pi}{4})\,
R_x^{(1)}(\frac{\pi}{2})\, R_x^{(2)}(\frac{\pi}{2})\, 
B(-\frac{\pi}{2},\frac{\pi}{2})\, R_x^{(1)}(\frac{3\pi}{4}) 
R_x^{(2)}(\frac{3\pi}{4}) R_y^{(2)}(\frac{\pi}{2}).
\label{eq:QFTflux}
\ee
${\cal F}_{\rm (SWAP)^{\al}}$ and ${\cal F}_{B}$ also give alternative implementations of 
the CNOT$^{(2,1)}$ gate, which consist of applying
the sequences (\ref{eq:QFTspin}) or (\ref{eq:QFTflux}) twice. These, however, are longer 
than the sequences obtained by optimizing CNOT directly [Eqns.~(\ref{eq:CNOTspin}) and
(\ref{eq:CNOTflux})],
since they do not implement the transformation from standard to maximally entangled basis 
with the least amount of two-qubit operations. 

\section{Application to electron spin qubits and superconducting flux qubits}
\label{sec-application}

For solid-state qubits, implementation of two-qubit quantum gates in an efficient way is 
important because of their short coherence times. In this section we discuss
the implementation of the above results for electron spin qubits and superconducting flux qubits 
and estimate relevant time scales. 
An electron spin qubit consists of a single electron confined in a quantum
dot, a small island in a semiconductor structure that can be filled with electrons in a controlled
way~\cite{elze05}. The qubit is encoded in the spin degree of freedom. Single qubit rotations 
can be generated by applying an electron spin resonance (ESR) pulse to the electron which is 
described by the evolution operator $U_R(t)= \exp [-(i/\hbar)\int_0^t {\mathcal H}_R(\tau) d\tau]$, with 
${\mathcal H}_R(t) = - \frac{1}{2}\hbar \gam \vec{B}(t)\cdot \vec{\sig}$. 
Rotations of qubit 1 and 2 are then represented as:
\be
R_n^{(1)}(\beta) = e^{-\frac{1}{2}i\beta \vec{n}\cdot \vec{\sig} \otimes I }\ \ \mbox{\rm and} \ \
R_n^{(2)}(\beta) = e^{-\frac{1}{2}i\beta I \otimes \vec{n}\cdot \vec{\sig} },
\label{eq:rotation}
\ee
with $\beta(t)$ $\equiv$ $-\gamma \int_0^{t} B(\tau) d\tau$, $\gam$ the gyromagnetic ratio,
$B(t)$ the magnetic ESR field (applied in the direction perpendicular to the Zeeman-splitting field), 
$\vec{n}$ $\equiv$ $(\sin \theta \cos \phi,\sin 
\theta \sin \phi,\cos\theta)$ a unit vector on the Bloch sphere [$\theta$$\in$$[0,\pi)$, 
$\phi$$\in$$[0,2\pi)$], $\vec{\sig}\equiv (\sig_x,\sig_y,\sig_z)$ and $I$ the identity matrix. 
Two-qubit interactions arise from the Heisenberg exchange interaction, so that 
the (SWAP)$^{\al}$ gate [Eq.~(\ref{eq:exch})] is the natural two-qubit gate for electron 
spin qubits. Both single-spin rotations and $\sqrt{\rm SWAP}$-operations have recently been 
demonstrated for spin qubits~\cite{kopp06,pett05}, and typical 
times are $t_{\rm rot} \sim 100$ ns (for magnetic fields of 1 mT) 
and $t_{\sqrt{\rm swap}} \sim 180$ ps.

Flux qubits~\cite{makh01} consist of a superconducting loop interrupted by three or four 
Josephson junctions. The qubit basis states consist of the direction of the current that is circulating
around the loop and single qubit rotations are generated by applying resonant microwave radiation. 
These rotations are described by the same expression~(\ref{eq:rotation}) as for spin qubits 
with $\beta \rightarrow
\frac{1}{\hbar} \int_0^t \Omega_j(\tau)\, \cos \phi_j(\tau) d\tau$ for $\sigma_x$ (rotations around
the $x$-axis) and $\beta \rightarrow - \frac{1}{\hbar} \int_0^t \Omega_j(\tau)\, \sin \phi_j(\tau) d\tau$ 
for $\sigma_y$ (rotations around the $y$-axis), $j=1,2$. Here $\Omega_j(\tau)$ and $\phi_j(\tau)$ denote
the amplitude and phase of the applied microwave signal. Measured Rabi oscillations are of the order of 
1-10 ns~\cite{chio03}. Various proposals to achieve a tunable coupling mechanism for flux qubits 
have recently been put forward~\cite{nisk06,graj06,plou04} and/or realized~\cite{hime06}. 
In each of these the proposed two-qubit 
gates are equivalent to the B-gate~\cite{plan07}. In Ref.~\cite{nisk06} the 
creation of controllable coupling between two detuned flux qubits via the quantum inductance of 
a third flux qubit is suggested, as described by the Hamiltonian~(\ref{eq:Hamflux}). 
Predicted B-gate operation times for this system range from 
$\sim 10$ ns~\cite{nisk06} down to $\sim 2$ ns~\cite{graj06}.

In practice, manipulating qubits introduces decoherence. The single-qubit decoherence 
time $T_2$ has not been measured yet for electron spins, but recent experiments show that 
single-spin Rabi oscillations remain visible for up to 1 $\mu$s (where each 
oscillation takes $\sim $ 100 ns)~\cite{kopp06} and ensemble decoherence times 
$T_2^{*} > 1 \mu$s~\cite{pett05}, so that $T_2$ is expected to be $\geq 1 \mu$s. 
For flux qubits, measurements of $T_2$ range from 15 ns~\cite{chio03} 
to a few $\mu$s\,~\cite{bert05}, while a Rabi oscillation requires $\sim 5$ ns. 
$T_2$ is thus sufficiently long to observe at least a few Rabi oscillations. It is unknown,
however, to what degree decoherence will affect a quantum gate operation which consists of several
single qubit rotations on different qubits as well as two-qubit interactions: minimizing the total number 
of operations used~\cite{grig05} is then likely to be an essential factor for achieving
high gate fidelities.

\section{Conclusion} 
\label{sec-conclusion}

We have presented a systematic protocol for calculating the optimal (defined as consisting 
of the smallest number 
of single- and two-qubit operations) implementation of a 
desired two-qubit gate $U\in SU(4)$ in terms of a given tunable two-qubit 
interaction $U_{int}(t)$ and single-qubit rotations. We have illustrated the decomposition method by 
calculating the shortest sequence of operations required to implement the transformation 
from the standard basis to the Bell basis, the CNOT gate, and the quantum Fourier transform, 
using either Heisenberg exchange interaction or quantum inductive coupling. 
The general method presented here is a useful tool to find the smallest number of operations 
that are needed to realize a desired two-qubit gate for any type of qubit, using the single- 
and tunable two-qubit operations that are naturally available in the qubit system.

\section{Acknowledgements}

This work has been supported by the Netherlands Organisation for
Scientific Research (NWO) and by the EU's Human Potential Research Network under
contract No. HPRN-CT-2002-00309 (``QUACS''). 

\section{Appendix}
\label{app_A}

\noi In this appendix we calculate the concurrence of the column vectors of the matrix
\be
T \equiv A(\eps) R_x^{(1)}(\mu) R_x^{(2)}(\nu) 
R_y^{(1)}(\xi) R_y^{(2)}(\zeta),
\label{eq:Tmatrix}
\ee
which is used in the proof of Proposition 1 in Sec.~\ref{subsec-max}. By writing out Eq.~(\ref{eq:Tmatrix})
explicitly, we obtain
\bea
T & = & \left( \ba{cccc}
B_1 & B_2 & -i B_2^{*} & i B_1^{*} \\
\al B_2 & \al B_1 & i\al B_1^{*} & -i \al B_2^{*} \\
-i \al B_2^{*} & i \al B_1^{*} & \al B_1 & \al B_2 \\
i B_1^{*} & -i B_2^{*} & B_2 & B_1 
\ea \right)
\left( \ba{cccc}
C_1 & -C_2 & -C_3 & C_4 \\
C_2 & C_1 & -C_4 & -C_3 \\
C_3 & -C_4 & C_1 & -C_2 \\
C_4 & C_3 & C_2 & C_1 \\
\ea \right) = \nn \\
& & 
\left( \ba{cccc}
B_1 C_1 + B_2C_2 -i B_2^{*} C_3 + i B_1^{*} C_4 & \ \ \ldots \ \ & \ \ \ldots \ \ & \ \ \ldots\ \ \\
\al (B_2 C_1 + B_1 C_2 + i B_1^{*} C_3 - i B_2^{*} C_4) & \ \ \ldots \ \ & \ \ \ldots \ \ & \ \ \ldots \ \ \\
\al (-i B_2^{*} C_1 + i B_1^{*} C_2 +  B_1 C_3 + B_2 C_4) & \ \ \ldots\ \ & \ \ \ldots \ \ & \ \ \ldots \ \ \\
i B_1^{*} C_1 - i B_2^{*} C_2 +  B_2 C_3 + B_1 C_4 & \ \ \ldots \ \ &\ \ \ldots \ \ & \ \ \ldots \ \
\ea \right),
\eea
with
\bea
B_1 & \equiv & \cos \frac{\mu}{2} \cos \frac{\nu}{2} - i \sin \frac{\mu}{2} \sin \frac{\nu}{2}
\nn \\
B_2 & \equiv & \sin \frac{\mu}{2} \cos \frac{\nu}{2} - i \cos \frac{\mu}{2} \sin \frac{\nu}{2}
\nn \\
C_1 & \equiv & \cos \frac{\xi}{2} \cos \frac{\zeta}{2} \nn \\
C_2 & \equiv & \cos \frac{\xi}{2} \sin \frac{\zeta}{2} \nn \\
C_3 & \equiv & \sin \frac{\xi}{2} \cos \frac{\zeta}{2} \nn \\
C_4 & \equiv & \sin \frac{\xi}{2} \sin \frac{\zeta}{2} \nn \\
\al & \equiv & \exp (-i \frac{\pi}{2} \eps) 
\eea
The concurrence $C$ of each of the column vectors $|w_i\ra$ of $T$ is found by calculating
the square roots $\lam_1 \geq \lam_2 \geq \lam_3 \geq \lam_4$ of the eigenvalues of the 
matrix 
\be
M \equiv \rho_i (\sig_y \otimes \sig_y) \rho_i^{*} (\sig_y \otimes \sig_y),
\label{eq:Mmatrix}
\ee
with 
$\rho_i \equiv |w_i\ra \la w_i|$, $i=1,\dots, 4$, and substituting
\be
C \equiv {\rm max} \{\lam_1 - \lam_2 - \lam_3 - \lam_4,0 \}.
\label{eq:concurrence}
\ee
The conditions on $\eps$, $\mu$, $\nu$, $\xi$ and $\zeta$ in Eq.~(\ref{eq:proposition}) are then 
obtained by evaluating $C=1$. The resulting expressions are in general lengthy and therefore not given here.
As an example, consider the special case $\xi$=$\zeta$=0. In this case, $T$ [Eq.~(\ref{eq:Tmatrix})] reduces to
\be
T_{\xi=\zeta=0} = \left( \ba{cccc}
B_1 & B_2 & -i B_2^{*} & i B_1^{*} \\
\al B_2 & \al B_1 & i\al B_1^{*} & -i \al B_2^{*} \\
-i \al B_2^{*} & i \al B_1^{*} & \al B_1 & \al B_2 \\
i B_1^{*} & -i B_2^{*} & B_2 & B_1 
\ea \right).
\label{eq:Tspecial}
\ee
We calculate the concurrence of the last column vector $|w_4\ra$ of Eq.~(\ref{eq:Tspecial}) (the
same result is obtained for the other column vectors). Let
\be
\rho = |w_4\ra \la w_4| = \frac{1}{2} \left( \ba{cccc}
|B_1|^2 & -\al^{*} B_1^{*} B_2 & i\al^{*} B_1^{*} B_2^{*} & i (B_1^{*})^2 \\
-\al B_1 B_2^{*} & |B_2|^2 & -i (B_2^{*})^2& -i\al B_1^{*} B_2^{*}  \\
-i \al B_1 B_2 &  i B_2^2 & |B_2|^2 & \al B_1^{*} B_2  \\
-i B_1^2 & i \al^{*} B_1 B_2 & \al^{*} B_1 B_2^{*} & |B_1|^2 
\ea \right).
\ee
Then $M$ [Eq.~(\ref{eq:Mmatrix})] is given by
\be
\frac{1}{2} \left( \ba{cccc}
|B_1|^2 P & - B_1^{*} B_2 Q & i B_1^{*} B_2^{*} Q & i (B_1^{*})^2 P \\
- B_1 B_2^{*} Q & |B_2|^2 R & -i (B_2^{*})^2 R& -i B_1^{*} B_2^{*} Q  \\
-i B_1 B_2 Q &  i B_2^2 R & |B_2|^2 R &  B_1^{*} B_2 Q \\
-i B_1^2 P & i  B_1 B_2 Q &  B_1 B_2^{*} Q & |B_1|^2  P
\ea \right),
\label{eq:Mspecial}
\ee
with 
\bea
P & \equiv & |B_1|^2 + (\al^{*})^2 |B_2|^2 \nn \\
Q & \equiv & \al |B_1|^2 + \al^{*} |B_2|^2 \nn \\
R & \equiv & \al^2 |B_1|^2 + |B_2|^2. \nn
\eea
The eigenvalues of Eq.~(\ref{eq:Mspecial}) are the solutions of the equation
\bea
& \lam^3 (\lam - 2 (|B_1|^2 P + |B_2|^2 R)) = 0 & \nn \\
& \Leftrightarrow  \lam = 0 \ \ \mbox{\rm or}\ \ \lam = |B_1|^4 + 2 \cos(\eps \pi)
|B_1|^2 |B_2|^2 + |B_2|^4 > 0
\eea
so that 
\be
C = \sqrt{|B_1|^4 + 2 \cos(\eps \pi) |B_1|^2 |B_2|^2 + |B_2|^4}.
\ee
Finally,
\be
C = 1\ \ \Leftrightarrow \ \ \left\{ \ba{l}
\eps = 0 \\
\eps \neq 0 \ \ \mbox{\rm and} \ \ (|B_1|^2 = 0, |B_2|^2 = 1 \ \ \mbox {\rm or}\ \ |B_1|^2 = 1, |B_2|^2 = 0).
\ea \right.
\label{eq:solution}
\ee
The latter solution corresponds to $\mu$, $\nu$ $\in$ $\{0, \pi, 2\pi, 3\pi, \ldots \}$.
Eq.~(\ref{eq:solution}) corresponds to the solution given in Eq.~(\ref{eq:proposition}).


\begin{thebibliography}{99}
\bibitem{similar}
A similar decomposition procedure has been given in Appendix A of Ref.~\cite{krau01}.
In contrast to the latter, however, we derive explicit expressions for the matrices
$K_1$, $A$ and $K_2$.
\bibitem{also}
Eqns.~(\ref{eq:lambdas}) were also given in Ref.~\cite{krau01}. In the following, we go beyond this result and derive
explicit expressions for $\al$, $\beta$ and $\gamma$.
\bibitem{others}
The values $\al$, $\beta$ and $\gamma$ given here (each of which depends on $\lam_1$-$\lam_4$) are
not unique and other expressions are equally valid, e.g. $\alpha =\frac{-i}{2}\log \left( \lambda _{1}\lambda _{2}
\right) +\frac{1}{2}k_{1}\pi$.
\bibitem{possible}
This is always possible and follows from Euler's theorem~\cite{bare95}.
\bibitem{zrotations}
If only rotations in the ($x$,$y$)-plane are available, $z$-rotations can be implemented
using $R_z^{(1)}(\phi)$ $=$ $R_y^{(1)}(-\frac{\pi}{2})$
$R_x^{(1)}(\phi)$  $R_y^{(1)}(\frac{\pi}{2})$ or $R_z^{(1)}(\phi)$ = $R_x^{(1)}(\frac{\pi}{2})$
$R_y^{(1)}(\phi)$  $R_x^{(1)}(-\frac{\pi}{2})$.
\bibitem{zdecomp}
This decomposition corresponds to the experimental set-up
in Ref.~\cite{kopp06} in which only rotations in the ($x,y$)-plane are available. An alternative implementation using $z$-rotations is
given in Ref.~\cite{burk99}, see Eq.~(8).
\bibitem{maxconc}
The maximum concurrence obtained when applying ${\cal F}$ to a separable 
quantum state $\in SU(2)\otimes SU(2)$ is 1/2.
\bibitem{swapQFT}
It is interesting to note that a ``swapped QFT'' can be implemented for
flux qubits using three operations less than the QFT itself: $({\rm SWAP} \cdot {\cal F})_{\rm flux}$ =
$R_y^{(1)}(-\frac{\pi}{2})$  $R_x^{(1)}(-\frac{3\pi}{4})$
 $B(\frac{\pi}{4},\frac{\pi}{4})$ 
$R_x^{(2)}(-\frac{3\pi}{4})$ $R_y^{(2)}(\frac{\pi}{2})$. When trying to reduce operational costs,
it is thus advantageous to use SWAP $\cdot {\cal F}$ instead of ${\cal F}$ in quantum protocols
for flux qubits.
\end{thebibliography}
\end{document}